\documentclass[amssymb,amsmath,aps]{revtex4}
\usepackage{natbib,graphicx,subfigure,subeqnarray,fancyhdr,amsmath,epstopdf,multirow,color, mathrsfs}

\begin{document}
\title{Les Houches 2012 Lecture Notes: \\
An introduction to the hydrodynamics of  locomotion on small scales}

\author{Eric Lauga}\email{e.lauga@damtp.cam.ac.uk}
\affiliation{Department of Applied Mathematics and Theoretical Physics, 
University of Cambridge, Cambridge CB3 0WA, United Kingdom.
}
\date{\today}

\begin{abstract}
I summarize below two presentations given at the {\it Les Houches} ``Soft Interfaces'' Summer School, July 2-27 2012, organized by 
L. Bocquet, D. Qu\'er\'e, and T. Witten.

\end{abstract}
\maketitle

\section{Introduction}
In these  lecture notes I will  briefly review the fundamental  physical principles of locomotion in fluids, with a particular emphasis on the low-Reynolds number world. The notes cover the material discussed in the three hours of my two lectures at the les Houches summer school, and as such cannot unfortunately do justice to the richness and variety of the field. A number of outstanding and longer pieces  should be consulted  by the interested reader. As a general introduction I recommend the wonderful book  {\it Life in Moving Fluids} by Vogel  \cite{vogel96}. A classic and early treatise on the different ways animals move is the book  {\it Animal Locomotion} by  Gray  \cite{gray68}.  A modern review for the hydrodynamic aspects at high-Reynolds number can be found in the comprehensive book  {\it NatureÕs Flyers: Birds, Insects, and the Biomechanics of Flight}  by Alexander  \cite{alexander02} while I recently co-authored a review article on aspects relevant to the low-Reynolds number world \cite{Lauga:2009ul}. An earlier comprehensive  article  focusing on the kinematics is that of Brennen \& Winet \cite{brennen77} while the one by Pedley \& Kessler addresses collective effects \cite{pedley92}. Readers specifically interested in the hydrodynamics of bacteria will find the book  {\it E. coli in Motion}  by Berg \cite{bergbook} the best entry point. Finally, for the mathematically-inclined, aspects of mathematical modeling are discussed at length by Lighthill in {\it Mathematical Biofluiddynamics} \cite{lighthill75}  and Childress in  {\it Mechanics of Swimming and Flying} \cite{childress81}.

\section{Locomotion in fluids}

\begin{figure}[t]
\begin{center}
\includegraphics[width=0.8\columnwidth]{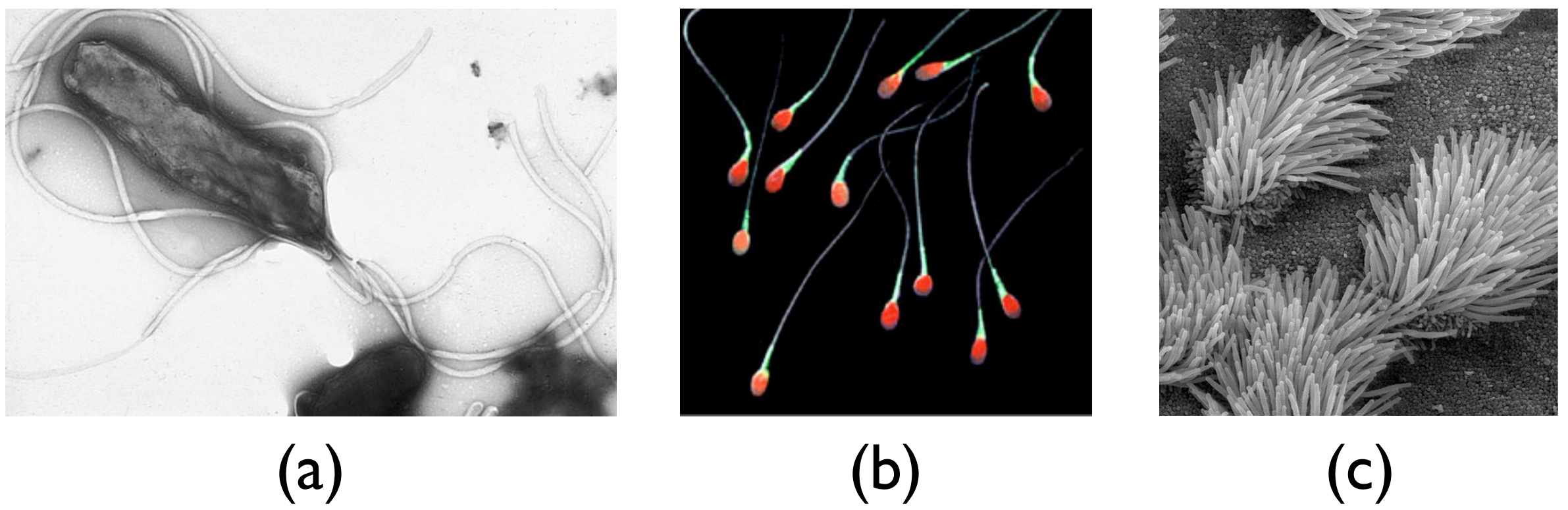}
\caption{Three examples of locomotion and transport on small scales. 
(a): Flagellated bacterium {\it Helicobacter pylori}; 
(b): Swimming human spermatozoa; 
(c): Cilia arrays on lung trachea epithelium. 
(Source: Wikimedia Commons).
}
\label{fig1}
\end{center}
\end{figure}

Cellular life on earth includes two different types of organisms: prokaryotes (bacteria and  archaea), and eukaryotes. Their distinction comes from the fact that prokaryotic cells do not possess a nucleus whereas eukaryotic cells do. In both worlds however, many organisms exploit the presence of a surrounding fluid for self-propulsion. In the world of  prokaryotes, many bacteria are motile. They actuate long and slender appendages, termed flagella, to swim in viscous and complex fluids (see \S\ref{real}). Examples of flagellated bacteria include {\it Escherichia coli}, {\it Bacillus subtilis}, or {\it Helicobacter pylori} (illustrated in Fig.~\ref{fig1}a). Other types of swimming bacteria have helical bodies such as the  spirochetes family or {\it spiroplasma}.

The domain of eukaryotic organisms includes four kingdoms (protists, animals, plants, and fungi), two of which include numerous examples of swimming in fluids. Protists include all single-celled protozoa,  many of which are swimmers, including {\it Paramecium} or {\it Euglena}. Algae also belong to the kingdom of protists, and  many planktonic organism show some form of propulsion, including the oft-studied single-celled algae {\it Chlamydomonas} and the multicellular {\it Volvox}. Beyond protists, the animal kingdom displays obviously a unique diversity in fluid-based locomotion. Without listing them all, one can mention jellies, worms,  flying  and hovering insects, mollusks such as gastropods crawling on mucus, and of course  vertebrates  including fish, amphibians, birds, seals, and  mammals. Within the animal kingdom, the swimming of spermatozoa during sexual reproduction  (Fig.~\ref{fig1}b) and the beating of lung cilia (Fig.~\ref{fig1}c) provide two relevant examples of locomotion and fluid transport  at low Reynolds number. 

In all these cases, locomotion in a fluid is achieved by the periodic  change of the shape of a particular organism. In a viscous fluid of density $\rho$ and dynamic viscosity $\eta$, shape changes lead to the generation of stresses in the fluid which propel the organism forward. When the time-averaged swimming  speed is denoted $U$, and the typical organism size is $L$, fluid-based locomotion ends up being characterized by a single dimensionless parameter, the Reynolds number $\Re= \rho U L/\eta$.

For the most part, locomotion in fluids in the biological or industrial world (man-made machines such as airplanes or submarines) occurs either at very large or at very small Reynolds number. Consider two illustrative examples. An olympic swimmer swims the 100 meter dash at a speed of about $U\approx 2 $~m/s. With a body size of $L\approx 2$~m and in water, this leads to a Reynolds number of $\Re\approx 4\times10^6$.  In contrast a small bacterium such as {\it E. coli} has  a length of about 2~$\mu$m, and swims with approximate speed  $U\approx 30$~$\mu$m/s, leading to a Reynolds number in water of about $\Re\approx 6\times10^{-5}$.

In these two lectures I focus on  biological fluid mechanics at small Reynolds numbers. My purpose is threefold. First I will emphasize the distinction between the low and the high Reynolds number world (\S\ref{Re-int}) and qualitatively explain the physical and mathematical consequences of swimming at low Reynolds number (\S\ref{2}). Second I will explain how real microorganisms are able to swim, show mathematically how to address their locomotion using  the canonical problem of the waving  locomotion of an inextensible flagellum, and show intuitively the role played by  noise on the locomotion of  the smallest cellular swimmers (\S\ref{real}, \ref{kinematics}, and \ref{diff}). Finally I will give a brief overview of what I think are interesting current  research questions  in the real of low-Re  locomotion (\S\ref{research}).

\section{Interpreting the Reynolds number in the context of locomotion}
\label{Re-int}

In order to shed light on the distinction between   the low-Re  world to its high-Re counterpart, let us consider the following  elementary calculation. Consider a swimmer of size $L$ steadily swimming in a viscous fluid (density $\rho$, viscosity $\eta$) with velocity $U$. At $t=0$ the swimmer instantaneously stops deforming  its shape periodically. How long does it take for the swimmer to come to a complete stop? Our intuition tells us that the resistance from  the surrounding environment will play a crucial role -- for example compare the time it takes for a bike to stop on a smooth road vs.~on mud.

The characteristic time $T$ necessary for the swimmer to stop is found by balancing, through Newton's second law, the drag from the surrounding fluid to the deceleration of the swimmer. Let us assume that the swimmer has density $\rho_s$ and thus its mass scales as $m\sim \rho_s L^3$. The deceleration from the swimmer is therefore on the order of $\sim \rho_s L^3 U/T$ where the value of $T $ is still to be determined. The scaling of the fluid drag, denoted $F_D$, depends critically on the typical value of the Reynolds number. If Re is large, $F_D$ has an inertial scaling, and we expect roughly $F_D\sim \rho U^2L^2 $. The balance between the fluid force and the deceleration leads to the simple estimate $ T \sim L   \rho_s /\rho U$. Given that the swimmer started at speed $U$,  the distance traveled  before coming to a complete stop, denoted $d$, is given by $d\sim UT \sim L \rho_s  /\rho$. Non-dimensionalized by the swimmer size we therefore see that 
this coasting length is given by the ratio of body to fluid density, $d/L\sim \rho_s /\rho$. For a human being at the swimming pool we thus get $d/L\sim 1$ whereas for birds in air we get $d/L\gg 1$, hence their ability to glide.

How does this scaling change in the low-Re world? When the Reynolds number is small, the fluid drag slowing down the swimmer has a different,  viscous, scaling $F_D\sim \eta L U$. As a result, the characteristic slowdown time now scales as $T \sim  \rho_s L^2 /\eta$, and thus the coasting length is now given by $d \sim \rho_s L^2 U /\eta$. If, as above, we nondimensionalize $d$ by the swimmer size $L$ we obtain $d/L \sim \rho_s L U /\eta$, which can be simply rewritten as
$d/L \sim\Re  \rho_s/\rho$ where $\Re$ is the Reynolds number based on the swimming speed. For microorganisms the ratio of density, $\rho_s/\rho$, is close to one, and  we finally obtain $d/L \sim\Re$. In the context of animal locomotion, the Reynolds number can thus be interpreted as a dimensionless gliding distance. 

Since for microorganisms the Reynolds number is typically much less than one, we thus obtain the result that  $d/L$ is essentially zero  \cite{purcell77}. The world of low-Re locomotion is therefore an instantaneous world, where organisms have to constantly input energy into the fluid. Physically, the forces acting on the microorganisms are dominated by viscous stresses from the fluid, and the inertial forces arising from the  velocity changes of swimmer can be neglected.  Living in a world subject to the laws of low-Re hydrodynamics leads to two notable physical consequences.

\section{The two physical consequences of low-Re swimming}
\label{2}

We saw in the previous section that microorganisms live in an instantaneous world governed by the physics of low-Reynolds number hydrodynamics. How can these physical constraints be exploited by organisms to generate self-propulsion?  This is done according to two physical principles.

\begin{figure}[t]
\begin{center}
\includegraphics[width=.4\columnwidth]{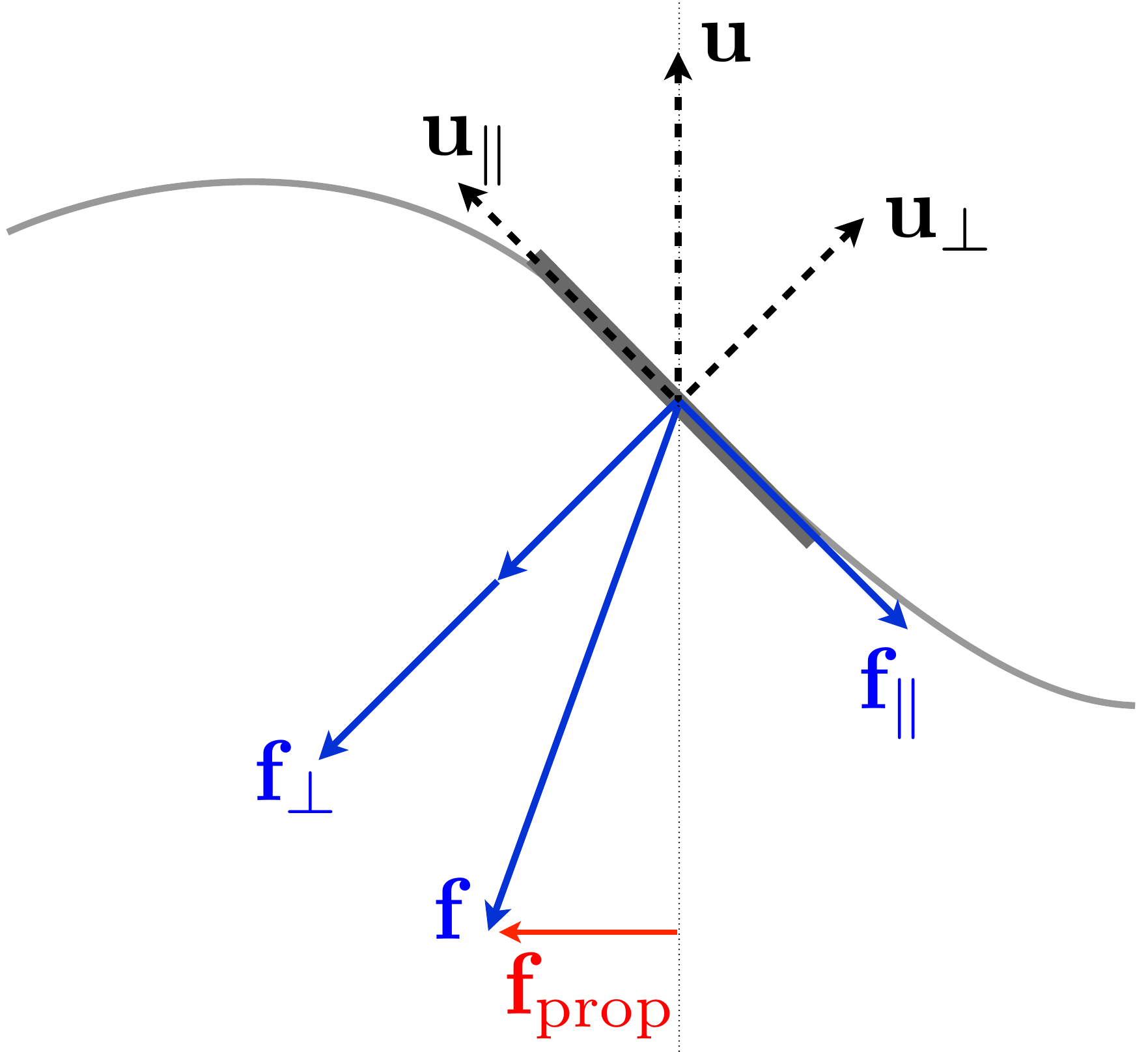}
\caption{Illustration of the principle of drag-based thrust along a flagellum. Because the drag coefficient for motion perpendicular to the flagellum is larger than that for motion parallel to it, the total drag force on the flagellum includes a nonzero propulsive component at right angle to the  local motion. 
}
\label{fig2}
\end{center}
\end{figure}

The first is that of drag-based thrust. In the macroscopic world we think of drag as something which impedes motion and thus it might sound somewhat counterintuitive that drag could be used to create motion. At low Reynolds number, the viscous drag on a moving body scales linearly with the speed of the body, but in general it does so in an anisotropic, tensorial, fashion -- in a manner which can be exploited to generate thrust. This physical principle is illustrated in Fig.~\ref{fig2}. Imagine zooming in on a beating flagellum. The filament moves relatively to the fluid with an instantaneous velocity denoted $\bf u$, which is typically perpendicular to the desired locomotion direction ($\bf u$ is vertical in Fig.~\ref{fig2} and the swimming direction is horizontal). The local fluid drag (per unit length) opposing the motion, denoted $\bf f$, scales linearly with the filament velocity in the form ${\bf f}=-{\bf C}\cdot{\bf u}$ where $\bf C$ is a resistance matrix. This linear dependence is a consequence of the linearity of the equations of fluid mechanics at low Reynolds number (see also below).  If the filament was spherical,  $\bf C$  would be an isotropic tensor, and the fluid drag would also be in the vertical direction. However, the filament is locally slender, and in that case $\bf C$  is  anisotropic. If we denote $\bf t$ the local tangent to the flagellum, $\bf C$   takes the  approximate form ${\bf C}\approx c_\parallel {\bf t \bf t} + c_\perp (\bf 1 - \bf t \bf t) $, where   
the ratio $c_\perp/c_\parallel > 1$ is  slightly less than 2 \cite{happel}. Drag on a slender filament translating along its length is thus about half the drag for the same filament translating perpendicularly to its length. 

As shown graphically in Fig.~\ref{fig2}, this non-isotropy of the drag is sufficient to generate a propulsive force at right angle with the flagellum motion. Decomposing the local filament velocity along its parallel (${\bf u}_\parallel=({\bf u}\cdot {\bf t}){\bf t}$) and perpendicular (${\bf u}_\perp={\bf u} - {\bf u}_\parallel$) components allows one to easily  compute  the parallel (${\bf f}_\parallel=-c_\parallel {\bf u}_\parallel$) and perpendicular (${\bf f}_\perp=-c_\perp {\bf u}_\perp$) components of the fluid drag and, by adding them up, the total fluid drag $\bf f$. As $c_\perp > c_\parallel$, the perpendicular force component is relatively larger than the parallel one, and the net force $\bf f$ is thus not exactly in the same direction as that of the filament velocity, $\bf u$. If we denote by $\hat{\bf u}$ the direction of the instantaneous velocity,  i.e.~$\hat{\bf u}\equiv {\bf u}/|{\bf u} |$, then we see that the propulsive force, ${\bf f}_{\rm prop}$, induced at right angle with the flagellum velocity is given by
\begin{equation}
{\bf f}_{\rm prop}= ({\bf 1} - \hat{\bf u}\hat{\bf u})\cdot {\bf f}=(c_\perp -c_\parallel)({\bf u}\cdot {\bf t})({\bf 1}-\hat{\bf u}\hat{\bf u})\cdot{\bf t}.
\end{equation}
If the swimming direction is  not perpendicular ($\hat{\bf u}\cdot{\bf t}\neq 0$)
or parallel ($\hat{\bf u}\cdot{\bf t}\neq 1$) to the local tangent to the filament, a nonzero propulsive force is thus induced  along the swimming direction (${\bf f}_{\rm prop}\neq {\bf 0}$). This is the physical principle for drag-induced thrust and one which is  exploited by the majority of swimming microorganisms.

In addition to taking advantage of fluid drag to generate propulsion, a second physical consequence of life at low Reynolds number has to be carefully considered. The description above of drag-induced thrust is a local statement, both local spatially along the flagellum and local in time (i.e. instantaneous). The propulsive force induced locally in Fig.~\ref{fig2} is pointing to the left, but perhaps somewhere else along the flagellum an identical force is being induced pointing this time to the right. In addition,    organisms undergo periodic shape changes, and thus perhaps at a later time in its periodic cycle,  the force at the location in Fig.~\ref{fig2} will point to the right. What is really important for propulsion is thus not only the instantaneous, local value of  ${\bf f}_{\rm prop}$, but for its  spatial and time average, denoted $\langle  {\bf f}_{\rm prop}\rangle$, to be nonzero. Instead of a local consequence, this is thus a global constraint on the distribution of the forcing in space and time at the whole-organism level.

The mathematical constraint associated with this observation is usually referred to as the scallop theorem, and was first introduced by E. Purcell in his famous lecture \cite{purcell77}. It is essentially  the theorem of kinematic reversibility for Stokes flows applied to locomotion  \cite{happel}. The equations governing the pressure, $p$, and velocity field, $\bf v$, are the incompressible Stokes equations ($\nabla p = \eta \nabla^2 {\bf v},\,\nabla \cdot {\bf v} =0$), which have the property of being linear and 
independent of time. The resulting locomotion speed, say $U$, of an organism changing its shape, $S$, in time is thus linearly proportional to the rate of change of the shape, with an expected scaling of the form $U\sim f(S) \dot S$, with $f$ a potentially complicated function of the instantaneous shape. If the sequence of shape is identical under a time-reversal symmetry, which is the case for example if it is described by a single degree of freedom, then the average locomotion speed, $\langle U \rangle$, is exactly zero. The essence of this constraint is simply that, in the absence of all inertial forces,  if the actuation on the fluid (swimming stroke) is time reversible, then the net locomotion gained during the first half of the stroke will be exactly canceled out during the second half on the stroke, independently of the rate at which either stroke is performed.

\section{How do real microorganisms swim?}
\label{real}

In order to be able to swim, microorganisms have to exploit the two constraints arising from  living in a low-Re world outlined above. They have to take advantage of drag in order to be able to generate thrust, and have to deform their body or appendages in a non-time-reversible manner. Combining the response to both constraints, microorganisms typically generate propulsion by actuating slender filaments, termed flagella or cilia, in a waving motion. A slender filament is able to generate a large ratio of drag coefficient, and thus exploit effectively drag-induced thrust. Deforming filaments in a waving fashion allows their kinematics to indicate a clear direction of time, and thus escape the global constraints of the scallop theorem. 

Prokaryotes and eukaryotes generate the  actuation  responsible for this filament motion in a different manner \cite{braybook}. Bacteria flagella are rigid helical filaments, which are passively actuated in rotation by rotary motors embedded inside the wall of the cells. In contrast, eukaryotic flagella are active filaments, which are internally actuated in bending by ATP-fueled molecular motors distributed all along their length, giving rise to waving deformation. As a consequence, whereas a bacteria flagellum has a fixed shape rotating in the fluid, an eukaryotic flagellum is constantly modifying its shape by balancing internal forcing, its passive elastic resistance, and the forces from the surrounding fluid.

\section{Example of kinematics: The waving motion of a flagellum}
\label{kinematics}
\begin{figure}[t]
\begin{center}
\includegraphics[width=.65\columnwidth]{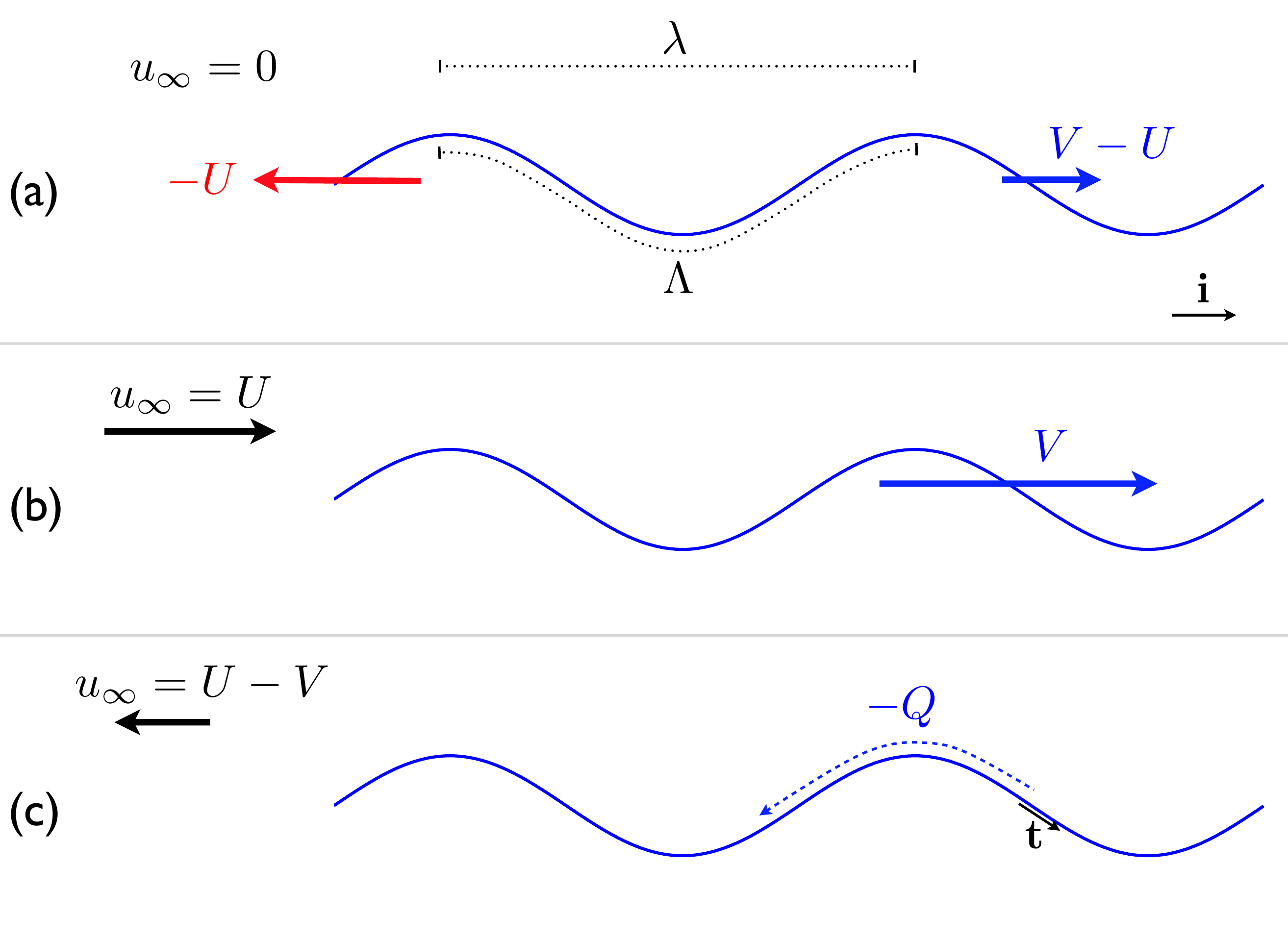}
\caption{Kinematics for the planar waving motion of an infinitely long, periodic flagellum. The flagellum swims to the left with  speed $U$ and deform its shape as a pure traveling wave moving with  speed $V$ to the right. The wavelength of the flagellum measured along its shape is $\Lambda$ while that measured along the swimmer axis (unit vector $\bf i$) is $\lambda$. 
Kinematics in the  laboratory frame (a), the  swimming frame (b) and the  
wave frame (c). The tangent vector to the flagellum shape is denoted $\bf t$ and the speed of the material points in the wave frame is $Q$.
}
\label{fig3}
\end{center}
\end{figure}

In order to gain insight into the relationship between the motion of an actuated flagellum and the resulting locomotion of a cell, we consider in this section a simple example, the waving motion of an infinitely long eukaryotic flagellum. This calculation is originally due to Lighthill \cite{lighthill75}, and we reproduce here its main steps  \cite{brennen77}. The equivalent classical calculation for a rotating helical flagellum is due to Chwang \& Wu \cite{chwang71}.

The setup is illustrated in Fig.~\ref{fig3}. The flagellum is assumed to have a fixed, periodic shape which is deforming as a traveling wave. Denoting by $\bf i$ the axis of the swimmer, and assuming the wave is top-down symmetric and propagates in the $+\bf i$ direction with wave speed $V$ in the swimming frame, we expect locomotion to occur in the $-\bf i$ direction, with unknown magnitude $U$.  In order to determine the value of $U$, we first need to determine the instantaneous velocity of points along the flagellum, then calculate the distribution of fluid forces. The swimming speed will then be the only value of $U$ leading to zero net force on the swimmer. For simplicity we neglect the presence of a head  \cite{brennen77}. 

The three panels of Fig.~\ref{fig3} allow to simplify the analysis in order to determine  the flagellum kinematics. In Fig.~\ref{fig3}a we illustrate the problem as it takes place in the laboratory frame, where the velocity at infinity, $u_\infty$, is zero. Since $V\bf i$ is the wave speed expressed in the moving frame, the apparent traveling speed of the wave in the laboratory frame is $(V-U)\bf i$. In Fig.~\ref{fig3}b we look at the same problem but expressed in the moving frame. The velocity at infinity is thus ${\bf u}_\infty=U \bf i$, and the wave is traveling with constant wave speed $V\bf i$. In order to be in a frame in which the shape of the flagellum is fixed we now have to jump into the  frame which is translating with the wave at speed $V\bf i$. The problem in that frame is summarized in Fig.~\ref{fig3}c. The speed at infinity is now ${\bf u}_\infty=(U -V)\bf i$ and, most importantly, in that frame  the shape of the flagellum remains constant.  Consequently, material points along the flagellum can only move tangentially to the flagellum, with speed denoted $- Q {\bf t}$,  where  $Q$ is  constant  to ensure inextensibility of the  flagellum. What is the value of $Q$? Let us denote by  $\lambda$ the wavelength of the periodic flagellum measured along the $\bf i$ direction and $\Lambda\geq \lambda$ the curvilinear  wavelength measured along the  flagellum itself (see Fig.~\ref{fig3}a). Geometrically, during one wave period,   wave crests  are displaced by an amount $\lambda$ along the $\bf i$ direction and  at speed $V$ whereas material points have to move a length scale of $\Lambda$ along the flagellum at speed $Q$, and thus we necessarily have $Q=({\Lambda}/{\lambda})V$.

Moving back into the laboratory frame, we finally get that each point moves with velocity 
\begin{equation}
{\bf u} = (V-U ){\bf i} - \frac{\Lambda}{\lambda}V {\bf t},
\end{equation}
and the spatial dependence of $\bf u$ occurs implicitly through the spatial variations of the tangent vector, $ {\bf t}$, to the flagellum. The dependence on the overall wave geometry comes from the  values of ${\Lambda}$ and ${\lambda}$. With the kinematics determined, we can use the results from  \S\ref{2} to compute the distribution of force per unit length, ${\bf f}$, due to the  surrounding fluid, ${\bf f}=-{\bf C}\cdot{\bf u}$. Solving for the value of $U$ which leads to a  net zero force on the swimmer, $\int {\bf f}{\rm d}s={\bf 0}$, leads to the solution for the swimming speed as
\begin{equation}\label{U}
\frac{U}{V}=\frac{(1-\gamma)(1-\beta)}{1+\beta(\gamma-1)},
\end{equation}
where $\gamma= {c_\parallel}/{c_\perp}$ is the ratio of drag coefficients (with a value slightly above  one half), and $\beta=[\int_0^\Lambda ({\rm d}x/{\rm d}s)^2{\rm d s}]/\Lambda\leq 1$ with $x(s)$ being the function describing the wave along the $\bf i$ vector ($s$ is the curvilinear coordinate along the flagellum).

The solution in Eq.~\eqref{U} is noteworthy for four reasons. First we see that $U>0$ and thus swimming occurs always in the direction opposite to that of the wave propagation. Secondly, it is clear that $U/V \leq 1$ and therefore the swimming speed is always below the wave speed. 
Thirdly, for isotropic drag, we have $\gamma=1$, and  thus $U=0$: drag anisotropy is therefore crucial to be able to swim.  Finally, the only geometrical parameter impacting the value of the swimming speed is $\beta$, which is the integral along one period of the square of the cosine of the angle between the local tangent and the swimming direction. 

To get an idea of the typical value expected for $U$, we can consider sinusoidal waves of the form $y(x) =a \sin (2\pi x/\lambda)$, where $a$ is half the peak-to-peak amplitude. Elementary algebra shows that in that case, $(dx/ds)^2=[1+(2\pi a/\lambda )^2\cos(2\pi x/\lambda)^2]^{-1}$ and thus
\begin{equation}
\beta = \frac{1}{\lambda}\int_0^\lambda \frac{{\rm d} x}{1+(2\pi a/\lambda )^2\cos(2\pi x/\lambda)^2}\cdot
\end{equation}
Assuming $\gamma =1/2$, and taking a wave with peak-to-peak amplitude equal to the wavelength ($a=\lambda/2$),  we obtain $\beta\approx 0.3 $ and $U/V\approx 0.4$. If instead the peak-to-peak amplitude is one third of the wavelength ($a=\lambda/6$), we get $\beta \approx0.7 $ leading to $U/V\approx 0.25$. 

\section{Locomotion vs.~diffusion}
\label{diff}

With our understanding of how microorganisms are able to generate the forces propelling them  in a viscous fluids, we are now able to address problems relevant to the interactions of swimmers with their environment. Some of these questions are active research topics and we will briefly overview them in \S\ref{research}. One particularly  fundamental issue concerns the role of noise and fluctuations in the cells dynamics.  The discussions in the previous sections were made under the assumption of deterministic fluid mechanics, with no consideration of noise. Noise can arise from a variety of sources. Thermal noise leads to fluctuations in flagella shapes, Brownian motions of the cells, as well as their  reorientation. Athermal noise can also arise from microscopic fluctuations in the behavior of molecular motors or by the   the random changes in the actuation at the whole cell level, in particular for bacteria \cite{bergbook_93}.  Physically,  the deterministic approach  outlined above will be valid on short time scales, whereas on longer time scales (to be defined) the effect of noise becomes important, and cells will typically undergo effective diffusion. 

\begin{figure}[t]
\begin{center}
\includegraphics[width=.8\columnwidth]{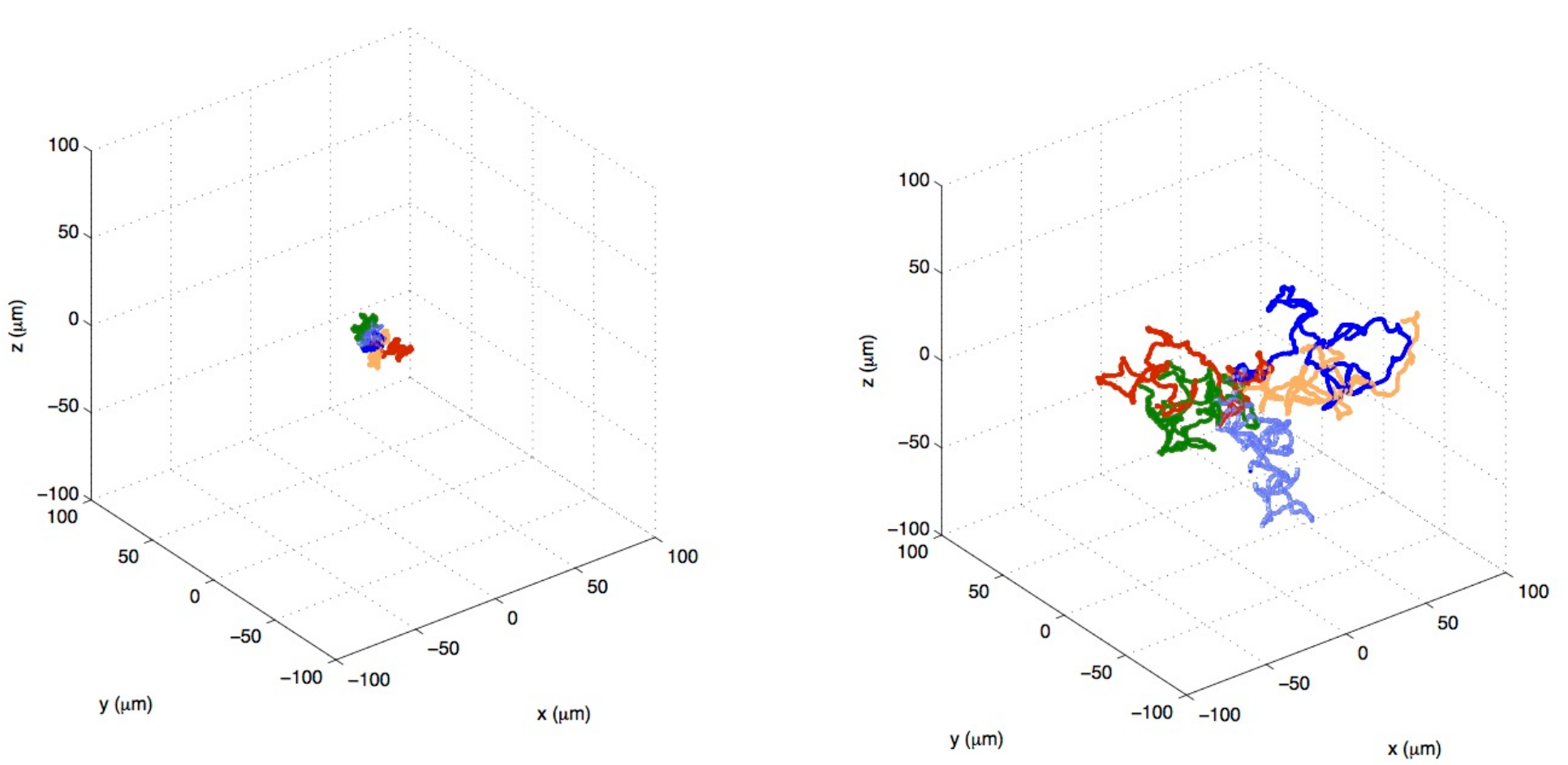}
\caption{
Effective diffusion of swimmers. 
We use Brownian Dynamics simulations on a spherical swimmer of 1 $\mu$m radius at room temperature to show that when the sphere is swimming (right, steady speed of 5 $\mu$m/s) it displays effective diffusive motion with diffusion constant much above that of purely Brownian motion (left, no swimming). Five different realizations are shown over a time scale of 100 s.
}
\label{fig4}
\end{center}
\end{figure}

What is  the purpose of bypassing the scallop theorem if, eventually, small swimming cells always end up showing zero time-averaged locomotion? The answer lies in the value of their diffusion constant. To illustrate this  point, let us consider the Brownian Dynamics simulations shown in in Fig.~\ref{fig4}. We show on the left the Brownian motion for a 1 $\mu$m-radius colloidal sphere in water and at room temperature over a time period of 100 s (5 different realizations are shown). On the right, in the same environment and over the same time period, we show the dynamics of the same spherical body in the case where it is able to swim at a speed of 5 $\mu$m/s. The combination of directed swimming with thermal reorientation of the sphere leads to an effective diffusion characterized by a diffusion constant significantly above that given by pure Brownian motion. Swimming does  not prevent cells from eventually diffusing, but it can allow  them to enhance their diffusion constants, possibly by orders of magnitude. 

A simple mathematical model allowing to quantify swimming-enhanced diffusion consists in approaching the cell dynamics as that of a random walk \cite{bergbook_93}. Imagine cells swimming along straight lines at a constant velocity $U$ for a time $\tau$ after which they reorient. This reorientation could be due to rotational diffusion, to a change in the swimming direction due to the `tumbling' process of bacteria with many flagella \cite{bergbook}, or another mechanism. The length of a step in the random walk is thus $d=U\tau.$ After a number $n$ of such steps, the mean square displacement of the cell would be $\langle x^2\rangle \sim n d^2$ and since the total time elapsed is $t=n\tau$, we obtain diffusion, $\langle x^2\rangle \sim D t$, with an effective diffusion constant scaling as $D\sim d^2/\tau \sim U^2\tau$.  

What is the critical reorientation time scale, $\tau_{\rm min}$, after which swimming always induce enhanced diffusion? It is found by setting the magnitude of $U^2\tau$ equal to the Brownian diffusion constant of the cell, $D_{k_BT}$, and thus we get $\tau_{\rm min} \sim D_{k_BT}/U^2$. For a 1~$\mu$m swimmer in water and room temperature we have $D_{k_B T}\approx 0.2$~($\mu$m)$^2$/s, and  a thermal reorientation time scale of a few seconds. With a swimming speed of  5~$\mu$m/s, this leads to $\tau_{\rm min} \approx 10$~ms, much less than the time scale for thermal loss of direction,  hence the very large enhancement in diffusion constant seen in Fig.~\ref{fig4}. Note that in the case where reorientation is due to diffusion in orientation only, the exact solution for the effective diffusion constant in three dimensions is $D=D_{k_BT} + U^2/6D_R$, where $D_R$ is the coefficient of rotational diffusion of the cell \cite{bergbook_93}.

Peritrichously flagellated bacteria such as {\it E. coli} have multiple flagella and they change their reorientation as a result of a so-called `tumbling' process during which at least one of the rotary motors driving a flagellum  changes its  rotation direction.    {\it E. coli} has a the cell body with a length scale  of about  $\approx$ 2~$\mu$m and swims at speed $U \approx$ 30~$\mu$m/s, leading to $\tau_{\rm min}$ on the order of 0.1~ms, which is  smaller than the typical time scale between reorientation events (on the order of 1 s) and again leads to enhanced diffusion  \cite{lovely75,bergbook_93}.  Note that other organisms, in particular marine bacteria,  employ reorientation mechanisms which have yet to be  fully elucidated  \cite{guasto12}. 

In his famous lecture, Purcell put forward a  physical argument giving an estimate of  the useful value of $\tau$ \cite{purcell77}. He argued that for bacteria such as {\it E. coli} which  swim in order to probe their chemical environment, the important thing is for cells to outrun the diffusion of nutrients -- allowing them to check whether indeed life was greener on the other side. On short time scales, the diffusive dynamics of a passive molecule will always be faster than the ballistic swimming of the cells, and thus swimming has to be sustained for a finite amount of time. Over a time scale $\tau$, a cell swimming straight explores an environment of size $\sim U \tau $ whereas a nutrient molecule characterized by a diffusion constant $D_0$ explores a typical size $\sim (D_0 \tau )^{1/2}$. Cells have thus to swim at least during a time such that $U \tau > (D_0 \tau )^{1/2}$, or $ \tau >\tau_c$ with $\tau_c= D_0/U^2 $. With a typical molecular  value  $D_0\approx10^{-9}$~m$^2$/s and $U\approx30$~$\mu$m/s this leads to $\tau_c\approx 1$~s, on the order of the reorientation  time scale seen experimentally.   Cells do not need to swim for longer than that because they have now acquired the useful information about the local chemical nature of the new environment.

\section{Research questions}
\label{research}

So far I have  presented a quick overview of some of the classical results in the hydrodynamics of swimming microorganisms. In this last section I will  highlight  actively-pursued research questions in the field, emphasizing three themes. This is obviously a  personal point of view, and one  shaped by my own  interests. The current literature in the field is vast and I have tried to give enough references to provide appropriate entry points for the interested reader. 

The first active research theme is that of locomotion in complex environments. The physical results introduced in  the previous sections focused on  swimmers in  an infinite, Newtonian fluid in the absence of boundaries. In numerous biologically-relevant situations, the fluids are non-Newtonian, and locomotion occurs under confinement, for example during spermatozoa transport in mammalian reproduction \cite{suarez06,fauci06}. 

The impact of non-Newtonian stresses on locomotion has recently been the center of numerous studies. The main question addressed by these studies concerns the difference in swimming behavior in a non-Newtonian environment as opposed to a Newtonian fluid, and whether locomotion is helped or reduced by the change of environment. Small-amplitude analysis in viscoelastic fluids first showed that locomotion speeds were always reduced compared to the Newtonian situation for planar \cite{lauga07} or helical waving actuation \cite{FuPowersWolgemuth2007}. Computations showed however that for large-amplitude waving motion, the result could be the opposite, and viscoelasticity could in fact enhance  swimming \cite{teran2010}. Even in cases where swimming speeds are reduced, swimming in  viscoelastic fluids is  more hydrodynamically efficient  \cite{lailai-pre,laipof1}. Furthermore, the nonlinearities intrinsic to non-Newtonian fluids can be exploited to generate novel modes of propulsion, ones which are otherwise inefficient in Newtonian fluids  \cite{lauga_life,pak12}. Locomotion in suspensions and heterogeneous fluids was also considered, and in this case locomotion is enhanced by the presence of a fluid microstructure \cite{leshansky09}. Recent work has further investigated locomotion in gels, detailing in particular the conditions under which  locomotion might be helped by the presence of an underlying elastic network \cite{fu10}.

A small number of experimental studies were also able to address the role of viscoelastic stresses. The nematode {\it C. elegans} was  shown to decrease its swimming speed in an elastic fluid without modifying its swimming gait  \cite{arratia2011}. The locomotion of a force-free helix, used as a model for  locomotion induced by helical flagella,  showed a transition from reduced to enhanced swimming with an increase in the helix amplitude
\cite{Liu2011}. Finally, flexible synthetic swimmers driven by external fields under planar actuation were shown to  go faster than in a Newtonian fluid \cite{laugazenit13}. The physical picture emerging is that of a kinematic-specific impact of elastic stresses in the fluid on the locomotion performance: certain modes of locomotion are negatively affected while others are enhanced, in a manner which future work will have to fully unravel.

Boundaries and confinement have also been shown to affect  the spatial distribution of swimming microorganisms, their swimming kinematics, and their ability to generate propulsive forces.  Swimming cells are attracted by boundaries, a classical result recently revisited  theoretically and experimentally   \cite{fauci95,berke08,drescher2011fluid}, and thus in a confined environment cells are expected to always be located near boundaries. 
The detailed of the  hydrodynamic description in this case was the focus of studies on spermatozoa \cite{smith2009human,shum2010modelling}, active filaments \cite{evans2010propulsion}, flagellated bacteria \cite{giacche2010hydrodynamic}, and model microorganisms \cite{saverio12}.  For swimmers with chiral shapes, such as helical bacteria, the presence of a  wall close to the organism leads to the generation of a surface-induced hydrodynamic torque perpendicular to the surface. The combination of swimming and the presence of an external torque leads to  circular motion of the microorganism, with a rotation direction being of opposite sign for  swimming near a rigid wall  \cite{lauga06} vs.~near a free surface \cite{di2011swimming}.

Our second issue of interest concerns the collective modes of locomotion of microorganisms. We refer to recent review articles for in-depth discussion of the topic  \cite{ActiveReview,koch}. As an organism is swimming, it sets up in its vicinity a flow field which then exerts stresses on nearby cells, possibly affecting their orientation and locomotion. A natural question to ask is therefore whether these interactions are able to generate nontrivial collective behavior. 

The classical theoretical approach looks at the dilute limit in which organisms are modeled as point-dipoles. These theories predict generic instabilities for uniform suspensions of pusher-type swimmers being propelled from the back by their flagella (such as flagellated bacteria; this is in contrast with puller swimmers such as the algae {\it Chlamydomonas} swimming flagella-first) \cite{simha02,saintillan07,saintillan08,HoheneggerShelley2010}. Other theoretical approaches have also been proposed deriving effective hydrodynamic equations based on the detailed modeling of interactions between swimmers   \cite{BaskaranMarchetti2009}.

Experiments and simulations allow  the behavior of cells and model swimmers to be probed  beyond the dilute regime. Early experiments focused on suspensions of swimming bacteria and demonstrated the occurrence of coherent structures of jets and swirls with speed and length scales significantly above that of individual organisms  \cite{wu00,dombrowski04,tuval05,sokolov07}.  Numerical computations were able to reproduce features of this nonlinear dynamics, either by simulating ensemble of  model  swimmers \cite{hernandez2005transport} or by developing  a coarse-grained continuum approach \cite{aranon07}. Computations were also able to show that the generic instability also occurs for puller swimmers in the semi-dilute regime  \cite{Art_squirmer}, allowed to investigate the role of  bounding walls of collective locomotion \cite{hernandez2009dynamics}, and study the difference between coherent structures  in three dimensions \cite{Ishikawa2008} and monolayers \cite{ishikawa08}. One aspect in particular where collective motion is bound to play an important role is that of   rheology \cite{hatwalne04}. A recent  series of experiments showed that  active fluids display strongly shear-dependent viscosity \cite{sokolov10,rafai10}, something which might in turn affect a number of biological and biomedical transport phenomena. Future work will also have to uncover the impact of these collective modes of locomotion on chemical transport within the surrounding fluid and cell-cell communication. 

The final topic I wish to emphasize takes us outside the purely  biological realm and into bioengineering. The modes of locomotion seen in biological organisms have inspired the community to develop a number of synthetic swimming devices in the lab \cite{abbot,ozin05,nelson10}. Similarly to flagellated organisms with helical flagella, rigid magnetic swimmers have been proposed \cite{GhoshFischer09,NelsonRev}. In this case, a magnetic body is typically attached to a rigid helix, and an actuation in rotation of the head by an external magnetic field leads to rotation of the helix, and propulsion. Alternatively, the swimmer might be composed of a straight flexible filament in lieu of a flagellum, and that filament acquires a helical shape upon rotation by the external field \cite{flexible2,OnshunSoftMatter}. Similarly, a flexible filament under an external planar actuation undergoes a planar waving motion and is also able to swim \cite{WigginsGoldstein,dreyfus05}.

Beyond modes of locomotion directly inspired by the biological world, a number of other methods have also been proposed and implemented to generate self-propulsion on the micron scale. Most notably, the last ten years have seen a lot of activity in chemical, or catalytic, swimming \cite{chemical1}.  An early experiment showed that janus Pt/Au metallic rods  were able to swim in aqueous  solutions of hydrogen peroxide \cite{paxton04}. Physically, this class of small-scale swimmers are catalysts for a specific chemical reaction on a portion of their surface, while the rest of their body is inert. The asymmetry in the location of the catalyst leads to an asymmetry in the concentration of reactants and products of the chemical reaction, and a net  self-propulsion through self-diffusiophoresis or, in the case where the chemical reaction leads to the production of ions and electrons,  self-electrophoresis \cite{golestanian05,golestanian07,howse07,chemical2,chemical3}. Other types of synthetic swimming methods have included the idea of taking propulsive of advantage of the presence of  nearby surfaces, either rigid  \cite{surface1,surface2,surface3} or deformable surface \cite{trouilloud08}, as well as exploiting  
external acoustic fields \cite{wang12}.

\section{Conclusion}

 Over the last ten years, our knowledge on the physics of small-scale swimming has grown tremendously, based in part on the joint efforts of many different research communities: soft matter physics, applied mathematics, biophysics, colloidal science, chemistry, and biological physics. I hope  that these two lectures, even though they provide only a glimpse of  this active research field, are able to serve as an adequate introduction to its richness.

\section*{Acknowledgements}

I want to thank the organizers of the summer school (L. Bocquet, D. Qu\'er\'e, T. Witten) for including me in their great program on Soft Interfaces. I thank Yi Man, Fran\c cois Nadal, On Shun Pak, and  Gregory Wagner for giving me feedback on  these notes. 
Funding by the US National Science Foundation is gratefully acknowledged. 

\bibliographystyle{unsrt}
\bibliography{bib.bib}
\end{document}